# Heat Transfer in Lattice BGK Modeled Fluid

Y. Chen, H. Ohashi and M. Akiyama
*Department of Quantum Engineering and Systems Science*
*Faculty of Engineering, University of Tokyo*

**Abstract**

The thermal lattice BGK model is a recently suggested numerical tool which aims to simulate thermohydrodynamic problems. We investigate the quality of the lattice BGK simulation by calculating temperature profiles in the Couette flow under different Eckert and Mach numbers. A revised lower order model is proposed and the higher order model is proved once again to be advantageous.

## 1 Introduction

The lattice BGK method can be viewed as the latest development of the lattice Boltzmann(LB) method[2], which is a derivation of the lattice gas automata(LGA) model[1] for the simulation of fluid dynamics. Such a development was achieved by introducing the elaborate Bhatnagar-Gross-Krook collision operator[6] into the lattice Boltzmann equation(LBE)[3][4][5]. Although this single time relaxation approximation(STRA) made on the collision term of the discrete kinetic equation looks like oversimplified, the lattice BGK models amazingly reproduced well the complexities of fluid flows[7][8][9].

Recently, lattice BGK models in which thermal effects are included were explored by several authors[10][11][13]. These models are established by treating the conservation of particle kinetic energy non-trivially, which can be realized with the use of multi-speed particle distributions. The employment of the composed lattice is basically required, e.g. Alexander[10] employed the composed hexagonal lattice in two dimensional space, while others used the composed square lattice in the $D$ dimensional space. One may further divide these thermal lattice BGK models into two classes: the lower order models(Alexander and Qian's models[11]) and the higher order model(Chen's model[13]). The word "order" here refers to the order of the flow speed $u$ in its expansion of the equilibrium particle distribution, which appears in the lattice BGK equation as,

$$N_{pki}(\vec{x}+\vec{c}_{pki},t+1)-N_{pki}(\vec{x},t)=-\frac{1}{\tau}(N_{pki}-N_{pki}^{[eq]})\,. \quad (1)$$

Here, $N_{pki}$ represents the particle distribution on the $i$th link of the $pk$ sub-lattice, $\vec{c}_{pki}$ is the link vector and therefore the vector of the particle flight velocity, and $\tau$ is the relaxation time period during which particle distributions would approach to their equilibrium values. The equilibrium particle distribution, $N_{pki}^{[eq]}$, is written in the form of low speed expansion up to the 2nd(or 3rd) order of $u$ for the lower order models and 4th order for the higher order model. Correspondingly, different levels of symmetries of the underlying lattice are required for these models, so that the momentum and heat flux of the modeled fluid can be expressed in an isotropic form in the macroscopic limit. Specifically, the $n$th rank particle velocity-moment tensor defined as

$$T^{(n)}_{pk\alpha\ldots\xi}=\sum_i \overbrace{c_{pki\alpha}\ldots c_{pki\xi}}^{\text{n components}}\ , \quad (2)$$

is required to have an isotropic form up to $n=4$ for the lower models and $n=6$ for the higher order model. The latter requirement would prevent using the composed hexagonal lattice in $2D$ space and the FCHC lattice in the $4D$ space, for that the highest rank of the isotropic velocity-moment tensor is four for both cases[14].

Needless to say the higher order thermal lattice BGK model is more accurate, which was proved by the numerical measurement of the decaying rates of flow kinetic energy under different Mach numbers[13]. In that case, the higher order model was shown to be free from the deviation effects caused by the nonlinear terms hidden in the r.h.s. of the momentum equation of the modeled fluid. In this study, however, we shall concentrate on the aspect of heat transfer modeling for the thermal lattice BGK models. The investigation will be carried out by measuring and comparing the temperature profiles in the Couette flows under different Eckert numbers and Mach numbers. Models based on the composed square lattice are used in all these calculations. The key issues of the thermal lattice BGK model will be briefed in the next section, and numerical experiments and their results will be described subsequently. The lower and higher order models used in this paper will be labeled as follows,

- $2D13VQ$: Qian's lower order model, based on the $2D$ 13-link composed square lattice.
- $2D13VC$: a revised lower order model, based on the $2D$ 13-link composed square lattice.
- $2D16V$ : Chen's higher order model, based on the $2D$ 16-link composed square lattice.

Hence the discussions are confined in two dimensional cases, though the extension to three dimensional cases is straightforward. Some concluding remarks will be given in the last section.

# 2 Thermal lattice BGK models

## 2.1 Coordinates and symmetries of the composed lattice

The coordinates of the base vectors of the square sub-lattices may be written as,

$$k(\overbrace{\underbrace{\pm1,\pm1,\ldots,\pm1}_{\text{p components}},0,\ldots,0,0}^{\text{D components in D dimensions}})\,, \qquad (3)$$

and its permutations. The number of non-zero components is $p$ so that the moduli of such a vector is $|\vec{c}_{pki}| = k\sqrt{p}$. It is important to know, in the process of hydrodynamic derivation, the expression of the velocity-moment tensor $T^{(n)}_{pk\alpha\ldots\xi}$ which consists of the $n$-product of base vectors. The odd rank tensors vanish naturally by the definition itself. The even rank tensors, specifically 2nd, 4th and 6th rank tensors, can be written as follows in the $D$ dimensional space,

$$\begin{aligned} T^{(2)}_{pk\alpha\beta} &= \vartheta_{pk}\delta_{\alpha\beta}\,, \\ T^{(4)}_{pk\alpha\beta\gamma\delta} &= \psi_{pk}\Upsilon_{\alpha\beta\gamma\delta} + \varphi_{pk}(\delta_{\alpha\beta}\delta_{\gamma\delta}+\delta_{\alpha\gamma}\delta_{\beta\delta}+\delta_{\alpha\delta}\delta_{\beta\gamma})\,, \\ T^{(6)}_{pk\alpha\beta\gamma\delta\zeta\xi} &= \Lambda_{pk}\Upsilon_{\alpha\beta\gamma\delta\zeta\xi} + \Omega_{pk}(\delta_{\alpha\beta}\Upsilon_{\gamma\delta\zeta\xi}+\ldots) + \Theta_{pk}(\delta_{\alpha\beta}T^{(4)}_{pk\gamma\delta\zeta\xi}+\ldots)\,. \end{aligned} \qquad (4)$$

Here, $\delta$ is the Kronecker tensor and $\Upsilon$ is its higher order version. The ellipsis "..." stands for terms which can be obtained by permuting the indices of the foregoing term. The numerical values of parameters, such as $\vartheta_{pk}$, $\psi_{pk}$, ..., $\Theta_{pk}$ are listed in references[14][12] for one, two and three dimensions. The way to make the macroscopic flux tensors isotropic is to tune the particle populations on different sub-lattices properly so that the sum of anisotropic parts, parts that are related to $\psi_{pk}$, $\Lambda_{pk}$ and $\Omega_{pk}$, vanishes as a total effect.

## 2.2 Equilibrium particle distribution

When the flow speed of the modeled fluid is controlled to be much smaller than the flight speed of particles, the local equilibrium particle distribution can be expanded around the uniform equilibrium state. Considering flexibilities in the residence and the number of particles, and the parity invariance of the square lattice, the low speed expansion is written as follows,

$$\begin{aligned} N^{[eq]}_{pki} = {} & A_{pk} + M_{pk}(c_{pki\alpha}u_\alpha) + G_{pk}u^2 + J_{pk}(c_{pki\alpha}u_\alpha)^2 + \\ & Q_{pk}(c_{pki\alpha}u_\alpha)u^2 + H_{pk}(c_{pki\alpha}u_\alpha)^3 + R_{pk}(c_{pki\alpha}u_\alpha)^2u^2 + S_{pk}u^4 + \mathcal{O}(u^5)\,. \end{aligned} \qquad (5)$$

For the lower order models, terms whose orders are higher than $u^2$ may be cut out, or one of the third order terms can be retained to make the energy equation more accurate on the Euler level[12]. Parameters of the expansion depend on the local conserved quantities, namely density $\rho$ and thermal energy $e$. This dependence can be written in the following form,

$$X_{pk} = \rho \sum_{l=0}^{2} x_{pkl} e^l \,. \tag{6}$$

Here, $X_{pk}$ may represent any one of $A_{pk}$, $M_{pk}$, ..., $S_{pk}$ so that $x_{pkl}$ is actually $a_{pkl}$, $m_{pkl}$, ..., $s_{pkl}$. Various constraints, which may ensure the definition of the conserved quantities, the vanishing of anisotropic parts in the macroscopic flux tensors, and the nonexistence of unphysical artifacts in the macro-dynamic equations, can be imposed on these parameters. The number of such constraints is usually smaller than that of the parameters, so that the specification of a thermal lattice BGK model always involves some arbitrariness[11][12]. This implies that some optional constraints may be employed either to minimize the anisotropic effects or to improve the accuracy of the model, which would be made clear in the following sections.

## 2.3 Macroscopic flux tensors and transport coefficients

In the hydrodynamic derivation for the lattice BGK models, the discrete kinetic equation shall be first Taylor expanded into a continuous form in the long wavelength and low frequency limits. The macro-dynamic equations for the conserved quantities can be obtained subsequently by using the multi-scale technique, which is a perturbative formulation method. The small quantity for perturbation is proportional to the local Knudsen number and is denoted as $\epsilon$. Correct forms of the macro-dynamic equations are ensured by correct expressions of the macroscopic flux tensors, which are again ensured by the aforementioned constraints imposed on parameters of the low speed expansion of the equilibrium particle distribution. The resulted momentum flux tensors on the Euler($\epsilon$) order and Navier-Stokes($\epsilon^2$) order are given as,

$$\Pi^{(0)}_{\alpha\beta} = \sum_{pki} N^{(0)}_{pki} c_{pki\alpha} c_{pki\beta} = \frac{2}{D}\rho e \delta_{\alpha\beta} + \rho u_\alpha u_\beta \,, \tag{7}$$

$$\begin{aligned} \Pi^{(1)}_{\alpha\beta} &= \left(1 - \frac{1}{2\tau}\right) \sum_{pki} N^{(1)}_{pki} c_{pki\alpha} c_{pki\beta} \\ &= -\frac{2}{D}\rho e \left(\tau - \frac{1}{2}\right) \left[ (\partial_\alpha u_\beta + \partial_\beta u_\alpha) - \frac{2}{D}(\partial_\gamma u_\gamma)\delta_{\alpha\beta} \right] . \end{aligned} \tag{8}$$

Here $N^{(0)}_{pki}$ is the equilibrium part and $N^{(1)}_{pki}$ is the nonequilibrium part of the particle distribution. Heat flux vectors may be expressed on different orders as follows,

$$q^{(0)}_\alpha = \frac{1}{2} \sum_{pki} N^{(0)}_{pki} |c_{pki} - u|^2 (c_{pki\alpha} - u_\alpha) = 0 \,, \tag{9}$$

$$\begin{aligned} q^{(1)}_\alpha &= \frac{1}{2}\left(1 - \frac{1}{2\tau}\right) \sum_{pki} N^{(1)}_{pki} |c_{pki} - u|^2 (c_{pki\alpha} - u_\alpha) \\ &= -\frac{2(D+2)}{D^2}\rho e \left(\tau - \frac{1}{2}\right) \partial_\alpha e \,. \end{aligned} \tag{10}$$

Both the transport coefficients and the state equation of the modeled fluid can be easily identified from these formulas, e.g. the shear viscosity and the heat conductivity read respectively as,

$$\mu = \frac{2}{D}\rho e (\tau - \frac{1}{2}) \,, \tag{11}$$

$$\kappa = \frac{(D+2)}{D}\rho e (\tau - \frac{1}{2}) \,. \tag{12}$$

Note that the accuracies of Eqs. (7), (8), (9) and (10) are different for models of different orders. For the lower order models, $\Pi^{(0)}_{\alpha\beta}$ and $q^{(0)}_{\alpha}$ are accurate up to $\epsilon u^3$ and $\epsilon u^2$ orders, and $\Pi^{(1)}_{\alpha\beta}$ and $q^{(1)}_{\alpha}$ up to $\epsilon^2 u^2$ and $\epsilon^2 u$ orders. For the higher order model, the accuracies of the corresponding quantities are upgraded to $\epsilon u^5$, $\epsilon u^4$, $\epsilon^2 u^4$ and $\epsilon^2 u^3$ orders.

# 3 Numerical investigations

## 3.1 Couette flow

The flow system is an extremely simple one, consisting of one moving boundary, one rest boundary, and the fluid layer in between, see Fig. 1. In the case that the two parallel walls have identical temperatures, the dissipative work of the viscous force will still lead to a parabolic temperature distribution inside the fluid layer. From the viewpoint of dimensional analysis, the heat transfer of such a system shall be governed by two dimensionless parameters, namely the Prandtl number and the Eckert number. As the Prandtl number of the lattice BGK modeled fluid has an invariable unit value[12], the Eckert number, defined as

$$E = \frac{U_1^2}{c_p(T_1 - T_0)}, \tag{13}$$

becomes a decisive parameter. The definition of the Eckert number tells that it is a measure of ratio between the heat due to friction and the heat conducted by temperature difference. Note that this number will become infinite when the temperatures of two parallel walls are the same.

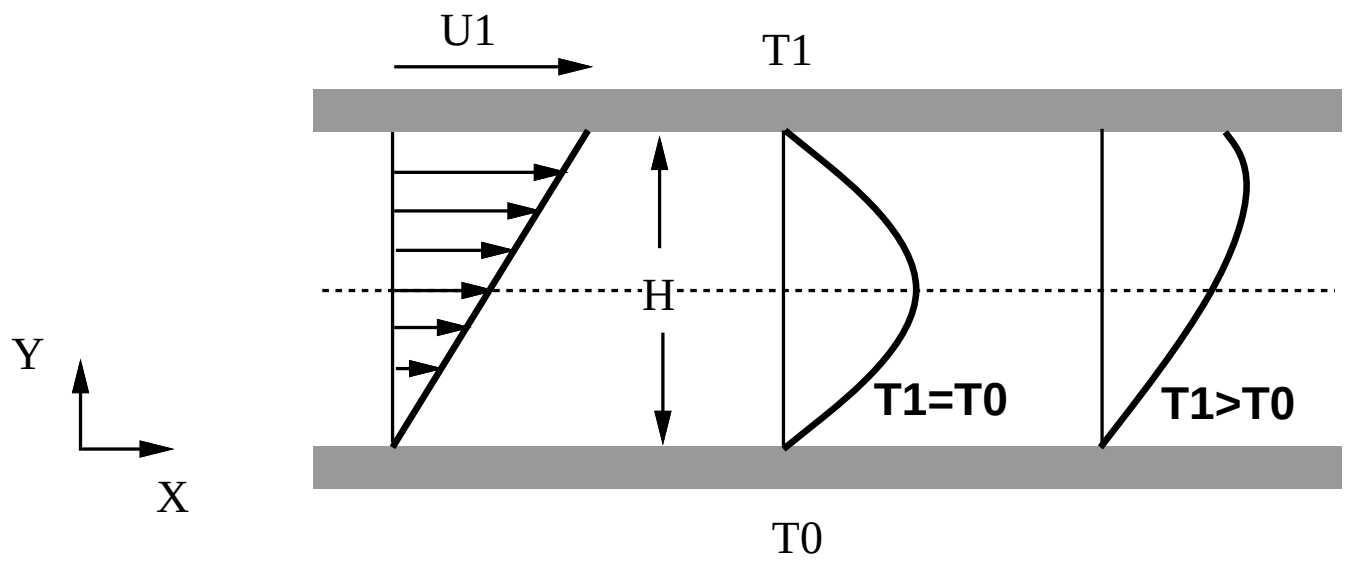

Figure 1: Velocity and temperature distribution in the Couette flow.

The analytical solution of the steady temperature distribution in the transverse direction of the channel can be obtained by directly solving the Navier-Stokes equations, which read as,

$$T_1 \neq T_0 : \quad T - T_0 = \frac{y}{H}(T_1 - T_0) + \frac{\mu U_1^2}{2\kappa}\frac{y}{H}\left(1 - \frac{y}{H}\right), \tag{14}$$

$$T_1 = T_0 : \quad T - T_0 = \frac{\mu U_1^2}{2\kappa}\frac{y}{H}\left(1 - \frac{y}{H}\right). \tag{15}$$

In the following calculations, numerical results will be normalized by $(T_1 - T_0)$ if $T_1 \neq T_0$ or by $\frac{\mu U_1^2}{2\kappa}$ otherwise. The linear distribution of the flow velocity will not be checked here, as it can always be obtained as long as the steady stage of flow is reached.

## 3.2 Results of Qian's model($2D13VQ$)

The model used here is a typical lower order thermal lattice BGK model on the composed square lattice. Conditions for the calculation are set as follows: lattice size $64 \times 32$, time step 10,000, moving boundary at $y = 1$ and non-slip and fixed temperature boundaries. The analytical solutions and the numerical results are shown in Fig. 2, from which one may conclude that the numerical calculation is accurate only when the Eckert number is small enough, that is, when the temperature distribution is mainly controlled by the heat conduction between the two walls. Notice that the Mach number $Ma = \frac{U_1}{a_s}$, where $a_s$ is the adiabatic sound speed, was kept to be smaller than 0.05 for all the three cases, in order that deviations caused by higher order terms contribute diminishing effects. It is clear, because of this low Mach number condition, that errors occurred under the large Eckert number are solely due to the lack of an exact expression for the viscous work in the energy equation.

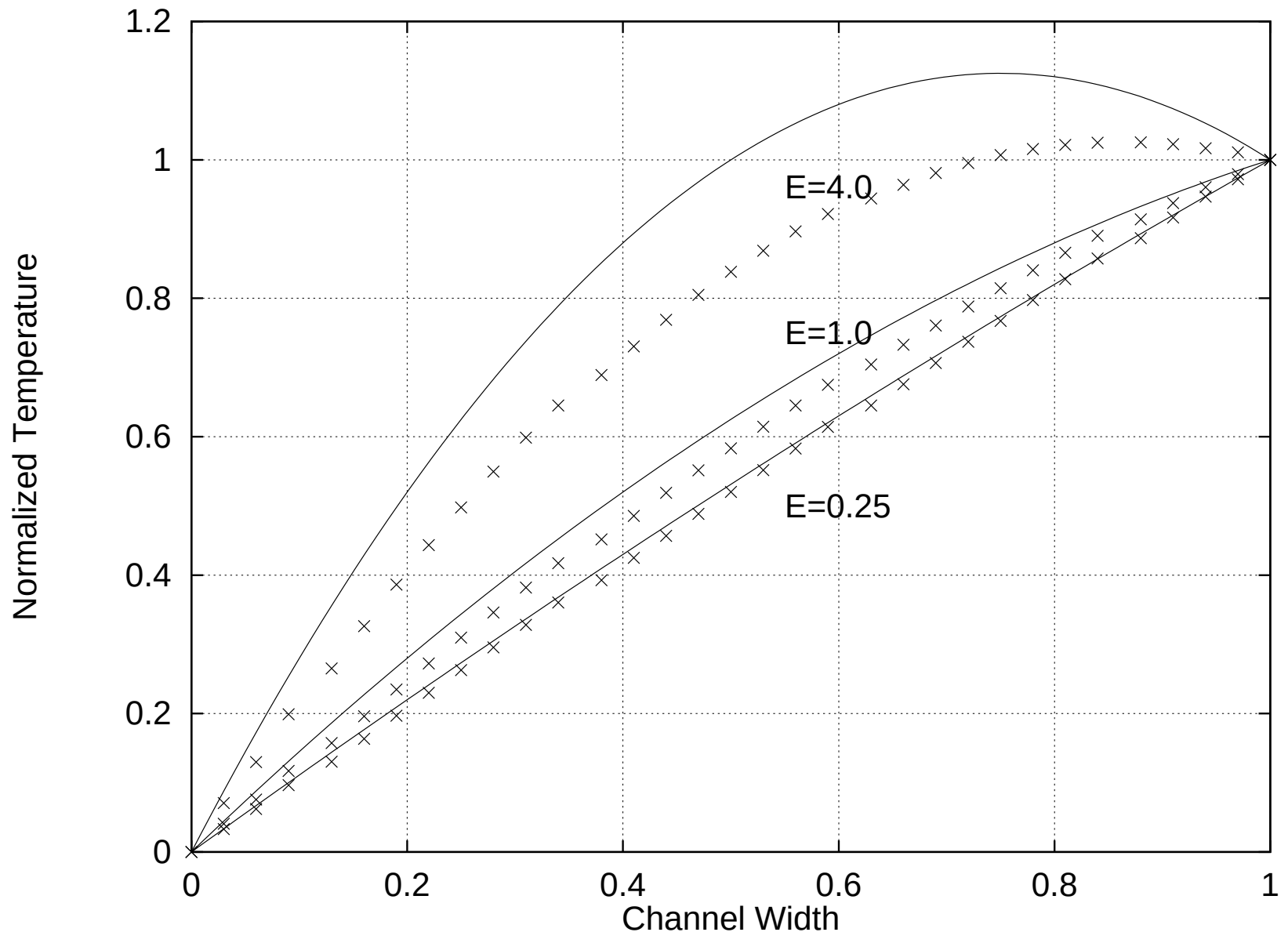

Figure 2: Temperature profiles in the Couette flow under different Eckert numbers. Solid lines are analytical solutions and symbols of crosses are the numerical results calculated by using Qian's thermal lattice BGK model.

It is crucial to derive the concrete expressions of the hidden terms in the energy equation of the modeled fluid, since one may then know the source of those large errors and improve the quality of lower order modeling if possible. We have derived the structures of these terms under two optional constraints,

$$\sum_{pk} \psi_{pk} g_{pk0} = 0 \,, \qquad \sum_{pk} \psi_{pk} g_{pk1} = 0 \,, \tag{16}$$

$$\sum_{pk} \Lambda_{pk} j_{pk0} = 0 \,, \qquad \sum_{pk} \Lambda_{pk} j_{pk1} = 0 \,. \tag{17}$$

As mentioned before, these optional constraints are used to minimize the anisotropic effects. The hidden terms can be explicitly written out in five parts[12],

$$\begin{aligned}
Part1 &: \frac{1}{2}\partial_\alpha(\rho u^2 u_\alpha) \,, \qquad (18)\\
Part2 &: \partial_\alpha[(\xi_1 + \xi_2\delta_{\alpha\beta})\partial_\beta(\rho e u_\alpha u_\beta) + (\xi_3 + \xi_4\delta_{\alpha\beta})\partial_\beta(\rho u_\alpha u_\beta)] \,,\\
Part3 &: -\partial_\alpha\left\{\left[\mu(\partial_\alpha u_\beta + \partial_\beta u_\alpha + \partial_\gamma u_\gamma\delta_{\alpha\beta}) + \sigma\left(1 - \frac{D-2}{2D}\delta_{\alpha\beta}\right)\partial_\gamma(\rho u_\alpha u_\beta u_\gamma)\right] u_\beta\right\}\\
Part4 &: \partial_\alpha\left\{\zeta_1\partial_\beta\left[\Delta^{[4,2]}\partial_\beta(\rho u_\delta u_\zeta)\right] + \zeta_2\partial_\beta\left[\Delta^{[4,2]}\partial_\beta(\rho e u_\delta u_\zeta)\right]\right\} \,,
\end{aligned}$$

$$Part5 \quad : \quad \sigma\left(1-\frac{1}{D}\delta_{\alpha\beta}\right)\partial_\gamma(\rho u_\alpha u_\beta u_\gamma)\partial_\alpha u_\beta\,.$$

Here, we use $\Delta^{[4,2]}$ to represent the sum of permuting products of the Kronecker tensors, namely $(\delta_{\alpha\beta}\Upsilon_{\gamma\delta\zeta\xi}+\ldots)$ appearing in Eq. (4). The nonlinear response coefficients are defined in the following formulas,

$$\begin{aligned}
\xi_1 &= \left[(D+4)\left(\sum_{pk}\theta_{pk}j_{pk1}\right)-\frac{D+2}{D}\right]\left(\tau-\frac{1}{2}\right), \qquad (19)\\
\xi_2 &= \frac{1}{2}\left[(D+2)\left(\sum_{pk}\varphi_{pk}g_{pk1}\right)+(D+4)\left(\sum_{pk}\theta_{pk}j_{pk1}\right)\right]\left(\tau-\frac{1}{2}\right),\\
\xi_3 &= (D+4)\left(\sum_{pk}\theta_{pk}j_{pk0}\right)\left(\tau-\frac{1}{2}\right),\\
\xi_4 &= \frac{1}{2}\left[(D+2)\left(\sum_{pk}\varphi_{pk}g_{pk0}\right)+(D+4)\left(\sum_{pk}\theta_{pk}j_{pk0}\right)\right]\left(\tau-\frac{1}{2}\right),\\
\zeta_1 &= \frac{1}{2}\left(\sum_{pk}\Omega_{pk}j_{pk0}\right)\left(\tau-\frac{1}{2}\right),\\
\zeta_2 &= \frac{1}{2}\left(\sum_{pk}\Omega_{pk}j_{pk1}\right)\left(\tau-\frac{1}{2}\right).
\end{aligned}$$

As the definitions of $\zeta_1$ and $\zeta_2$ involve $\Omega_{pk}$, the contribution of $Part4$ is anisotropic. Observing other parts, we see that there always exist terms of $\partial u^2$ order except in $Part1$ and $Part5$, which resulted from the higher order terms in the Euler level energy equation and the Navier-Stokes level momentum equation. Since the $\partial u^2$ terms have the same order as the normal viscous work terms, their combined influence should be the direct source of errors under the large Eckert number. Such errors, as discussed so far, are independent of the Mach number.

## 3.3 Results of a revised lower order model($2D13VC$)

The quality of the lower order thermal lattice BGK model can be improved by making changes in the optional constraints stated in the previous section. Actually we may relax one of such constraints, Eq. (16), and impose two new constraints as follows

$$\sum_{pk}\varphi_{pk}g_{pk0}=0\,, \qquad \sum_{pk}\varphi_{pk}g_{pk1}=-\frac{1}{D}\,, \qquad (20)$$

$$\sum_{pk}\Theta_{pk}j_{pk0}=0\,, \qquad \sum_{pk}\Theta_{pk}j_{pk1}=\frac{1}{D}\,. \qquad (21)$$

With the use of the definition of the shear viscosity in Eq. (11), it can be shown that these constraints will lead to the cancellation of terms of $\partial u^2$ order appearing in $Part2$ and $Part3$ of Eqs. (18). Nevertheless the relaxation of Eq. (16) will bring about an additional part of anisotropic errors, which involves the lattice symmetric parameter $\Lambda_{pk}$ and the higher order Kronecker tensor $\Upsilon_{\alpha\beta\gamma\delta\zeta\xi}$. Note that the anisotropic errors did not occur in the previous calculation, because $\Omega_{pk}$s' are zero in the two dimensional space. However, $\Lambda_{pk}$s' are non-zero parameters in all the dimensions, so that anisotropic errors will play their roles in the numerical calculation using the revised lower order model.

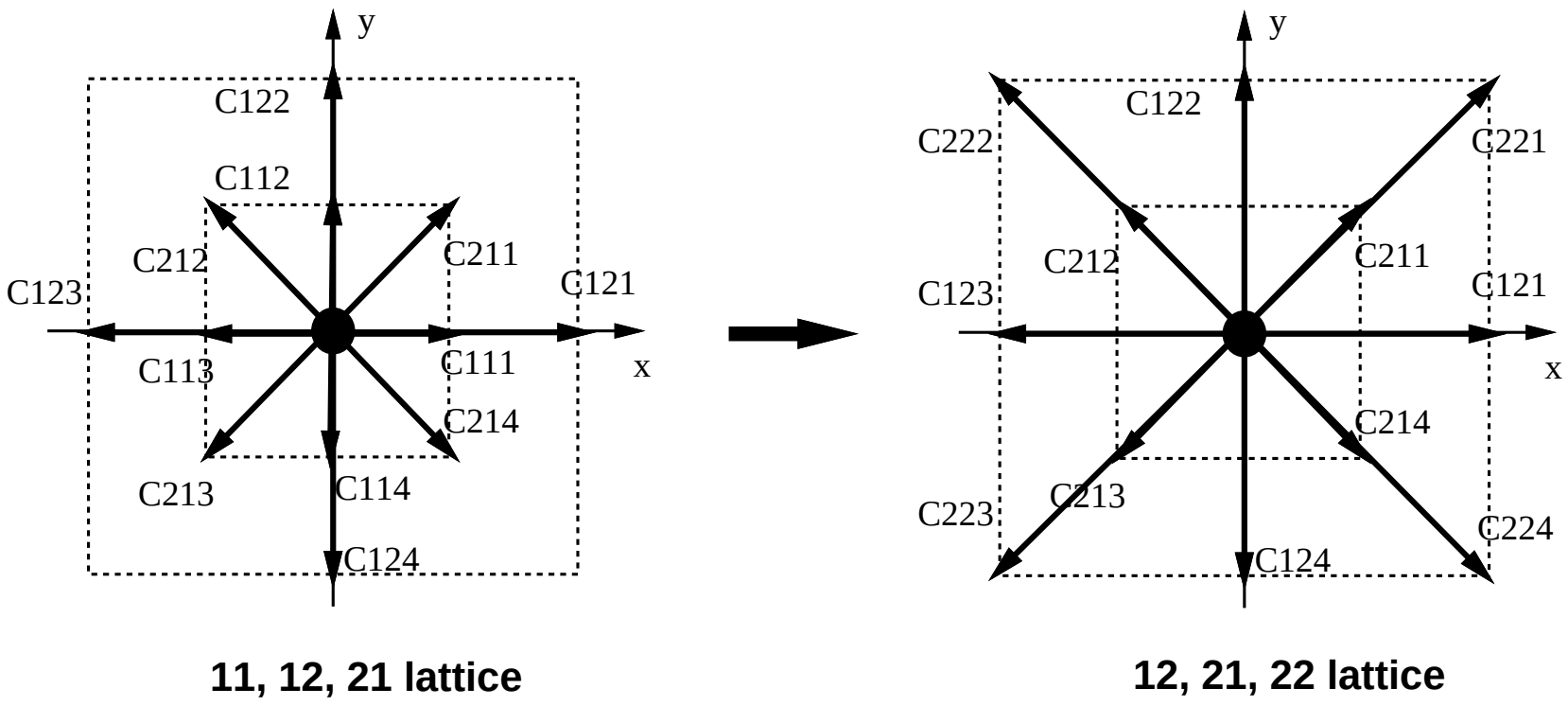

Figure 3: Change of the underlying lattice in the revised lower order model. The $2D13VQ$ model uses 00, 11, 12 and 21 sub-lattices and the $2D13VC$ model uses 00, 12, 21 and 22 sub-lattices.

The satifaction of Eq. (21), may not be realized without a further consideration of the lattice geometry. Since 00, 11, 12 and 21 sub-lattices are employed in the $2D13VQ$ model, $\Theta_{21}$ becomes the only non-zero parameter in the 6th rank velocity-moment tensor, and so is the $\varphi_{21}$ in the 4th rank tensor. Now that $j_{210}$ and $j_{211}$ would have already been decided by the $\varphi_{pk}$s' constraints, there would be no room for them to satisfy those $\Theta_{pk}$s' constraints. The solution is to replace the

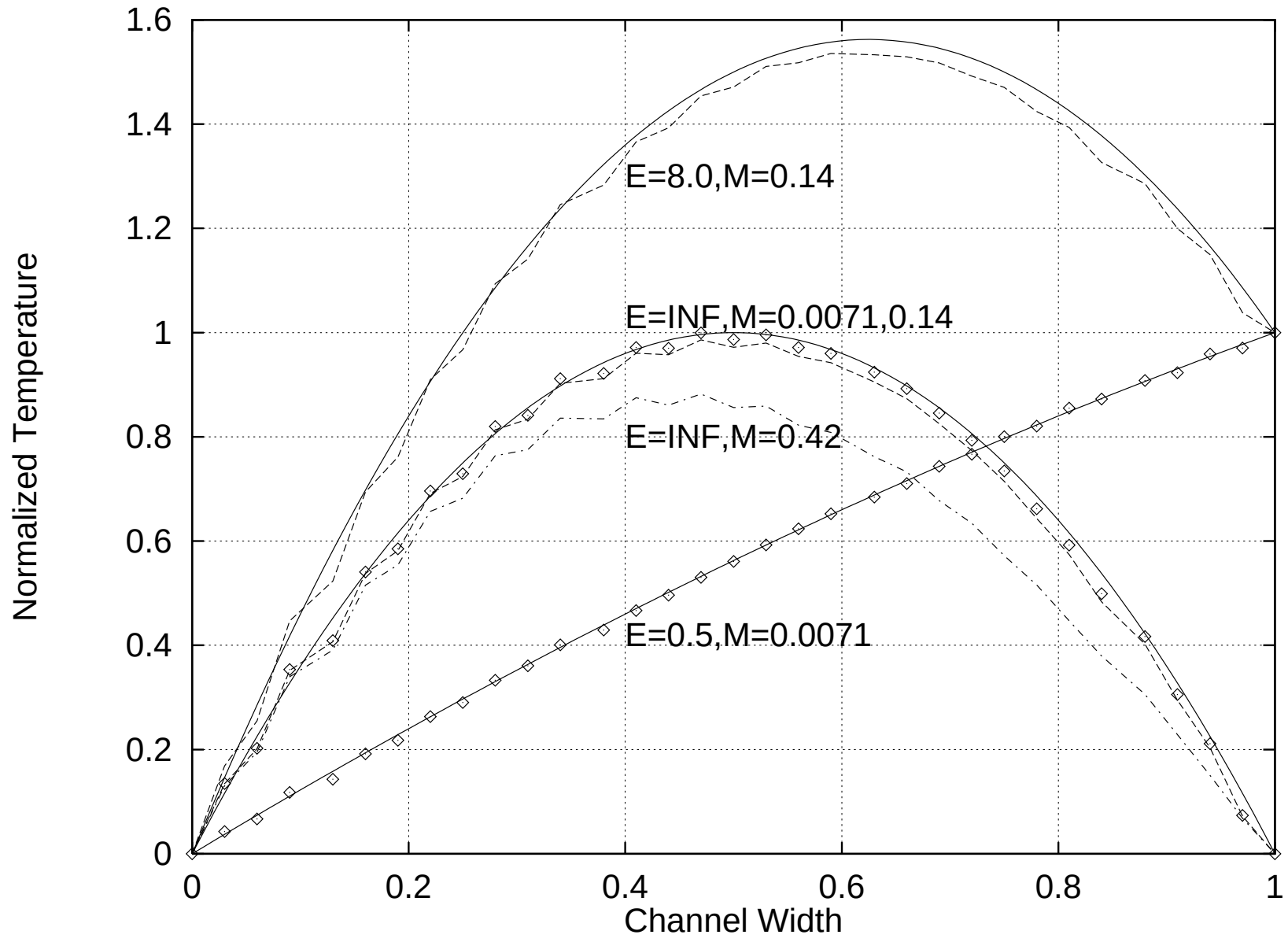

Figure 4: Temperature profiles in the Couette flow obtained by using the revised lower order thermal lattice BGK models. Solid lines are analytical solutions, others are the results of numerical calculations.

11 or 12 sub-lattice with the 22 sub-lattice. Parameters such as $\Theta_{21}$ and $\Theta_{22}$, $\varphi_{21}$ and $\varphi_{22}$ all have non-zero values, so that the satisfaction of both the $\varphi_{pk}$s' and $\Theta_{pk}$s' constraints becomes possible. We employed 00, 12, 21, 22 sub-lattices for the $2D13VC$ model, as shown in Fig. 3, taking care of the numerical stability. If the 11 sub-lattice is used instead, the difference of particle populations on different sub-lattice will be so large that particles may get nearly depleted on some links of certain sub-lattice. In that case, even a very small fluctuation could cause the appearance of negative particle distributions and thus the numerical instability is triggered on in the calculation.

Calculations were performed under the conditions described before. The results of the $2D13VC$ model and the analytical solutions are compared with each other in Fig. 4. It is found that errors have been greatly reduced even when the Eckert number becomes infinite. The remaining errors can be recognized as two parts. The first part of errors is brought by terms of $u^3$ or higher order. These errors become obvious when the Mach number is too large, see the broken line in Fig. 4. The loss of the symmetry for this line is due to the different local Mach numbers distributed across the section. Another part of errors comes from the anisotropic terms. The behavior of these errors in this case is an oscillating one, but small in both scale and magnitude. We are aware that even the anisotropic errors can be completely deleted if both the 11 and 12 sub-lattices are employed in the $2D13VC$ model. But this could make the lower order model cost the same amount of the computer resource as the higher order model, which is absolutely unfavorable, especially in the three dimensional case where the lower order model could save nearly half of the computer memory consumed by the higher order one.

### 3.4 Results of the higher order model($2D16V$)

It is clear now that the lower order thermal lattice BGK models suffer from different errors under different conditions. Although they are improvable, errors cannot be completely avoided. On the other hand, the higher order model was designed to eliminate all these deviations. It was proved before[13] that the deviation terms existing in the macroscopic momentum equation was removed for the higher order model . The calculation carried out here further shows that the modeling of heat transfer by using such a model is also advantageous. From the comparison of the numerical results and the analytical solutions shown in Fig. 5, we may know that the higher order model behaves extremely well no matter how the Eckert or the Mach numbers change.

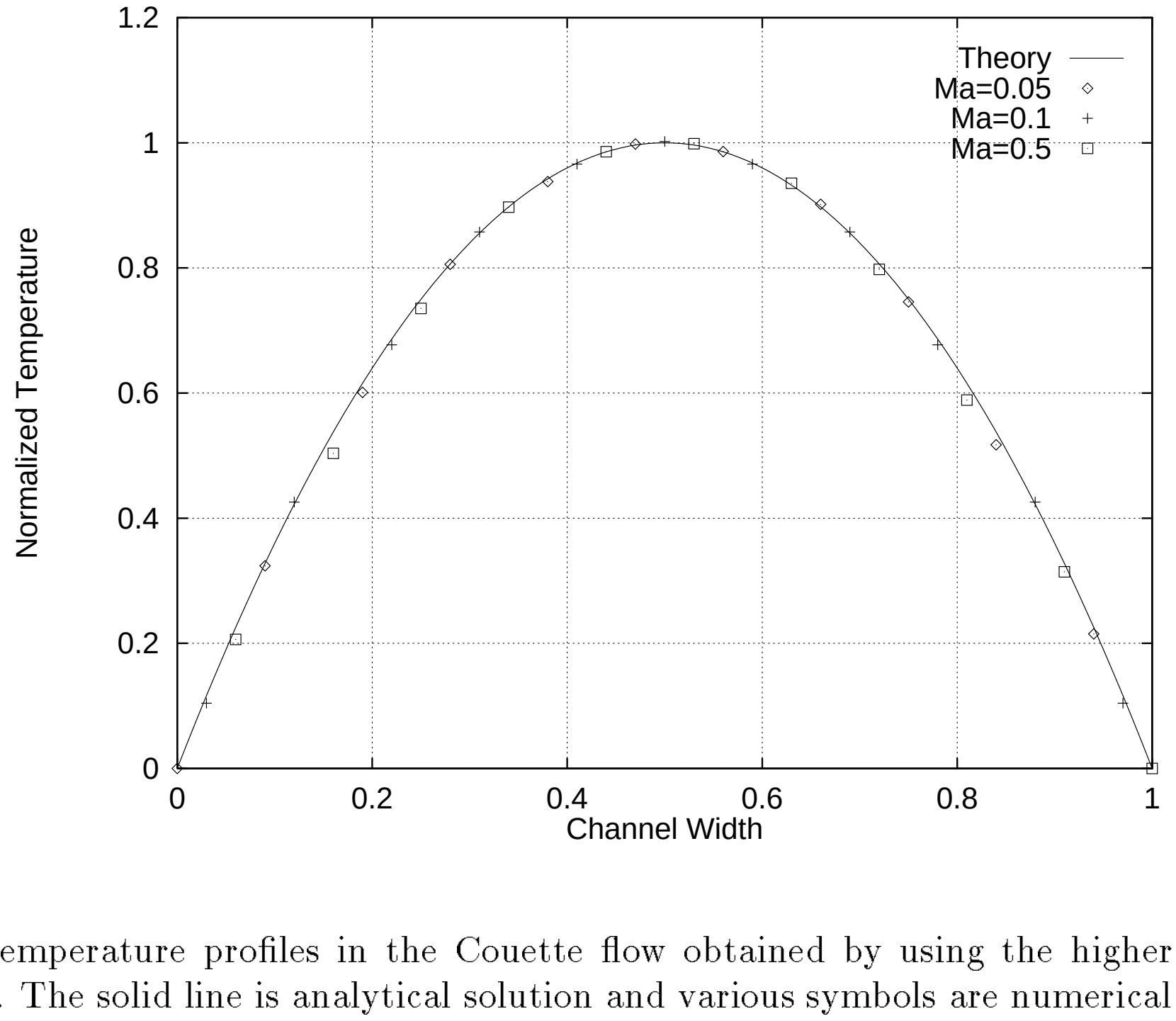

Figure 5: Temperature profiles in the Couette flow obtained by using the higher order lattice BGK model. The solid line is analytical solution and various symbols are numerical results under different Mach numbers.

## 4 Conclusion remarks

Through theoretical and numerical analyses presented above, the ability of the thermal lattice BGK models in modeling heat transfer in fluid flows was well demonstrated. Furthermore, we may contribute some advice to the users of such models: If one wants to save computer resource, specially in those $3D$ situations, and does not care about small scale anisotropic errors, it is safe to use the lower order thermal lattice BGK models in the incompressible regime. On the other hand, if high accuracies are required, then the higher order model is preferable. When flows enter the transonic regime, it is essential to use the higher order model to avoid large deviations in the simulation results.